\documentclass[12pt,aps,floats]{revtex4}
\usepackage{graphicx}
\usepackage{epsfig}
\usepackage{amssymb}
\usepackage{epstopdf}
\usepackage{amsmath}
\usepackage{epstopdf}
\usepackage{amsmath}
\usepackage{float}

\newskip\humongous \humongous=0pt plus 1000pt minus 1000pt

\relax

\begin{document}

\title{Remark on Remnant and Residue Entropy with GUP}
\bigskip
\author{Hui-Ling Li}
\email{LHL51759@126.com} \affiliation{College of Physics Science and Technology,
Shenyang Normal University, Shenyang 110034, China}

\author{Wei Li}
\affiliation{College of Physics Science and Technology,
Shenyang Normal University, Shenyang 110034, China}

\author{Yi-Wen Han}
\affiliation{School of Computer Science and Information Engineering, Chongqing Technology and Business University, Chongqing 400070, China}

\begin{abstract}
\textbf{Abstract:}

In this article, close to the Planck scale, we discuss on the remnant and residue entropy from a Rutz-Schwarzschild black hole in the frame of Finsler geometry. Employing the corrected Hamilton-Jacobi equation, the tunneling radiation of a scalar particle is presented, and the revised tunneling temperature and revised entropy are also found. Taking into account generalized uncertainty principle (GUP), we analyze the remnant stability and residue entropy based on thermodynamic phase transition. In addition, the effects of the Finsler perturbation parameter, GUP parameter and angular momentum parameter on remnant and residual entropy are also discussed.


\end{abstract}

\maketitle

\section{Introduction}
Although a complete self-consistent theory of quantum gravity has not been established, it is an effective way to understand the behavior of gravity by combining various models of quantum gravitational effects. In various quantum gravitation models, such as string theory, loop quantum gravity and non-commutative geometry, it is believed that there exists a minimum observable length, and this minimum observable length should have the order of the Planck scale. Considering a minimum observable length, a Hilbert space representation of quantum mechanics has been formulated by Kempf et al. in Ref. [1], which plays an important role on modifications of general relativity and black hole physics for research programs in recent years. Furthermore, Nozari and Etemad generalized the seminal work of Kempf et al. to the case that there is also a maximal particles' momentum, which resolved some shortcomings such as an infinite amount of energy for a free test particle, and provided several novel and interesting features \cite{2}. Especially, in very recently, Nozari et al. have addressed the origin of natural cutoff, including a minimal measurable length, and suggested that quantum gravity cutoffs are global (topological) properties of the symplectic manifolds \cite{3}. Now, one of the research directions of quantum gravity is to construct new theoretical models from the minimum observable scale, including Double Special Relativity (DSR) \cite{4}, Gravity's Rainbow \cite{5,6}, Modified Dispersion Relation (MDR) \cite{7} and Generalized Uncertainty Principle (GUP) \cite{1,8,9,10} and so on. Due to the problems of semi-classical tunneling radiation, the quantum gravitational effect caused by GUP is considered to study the quantum tunneling radiation of black holes£¬which leads to a very interesting result: GUP can prevent a black hole from evaporating completely, leaving remnant.

Recently, based on GUP, people began to study the quantum tunneling and remnant of black hole. Combining the GUP with the minimum observed length into the tunneling method of Parikh and Wilczek, Nozari and Mehdipour \cite{11} have successfully derived the quantum tunneling radiation of scalar particles from a Schwarzschild black hole. They pointed out that GUP has an important correction to the final state of black hole evaporation, and the black hole cannot evaporate completely. Following that, the possible effects of natural cutoffs as a minimal length, a maximal momentum, and a minimal momentum on the quantum tunneling have been completely discussed \cite{12}. Later, by utilizing the deformed Hamilton-Jacobi equations, Benrong et al. pointed out that breaking covariance in quantum gravity effective models is a key for a black hole to have the remnant left in the evaporation \cite{13}. In addition, in the frame of improved exponential GUP \cite{14}, the remnant of a Schwarzschild black hole has been given at the end of the evaporation process \cite{15}. On the other hand, considering the quantum gravitational effect under the influence of GUP, adopting the fermion tunneling method and the modified Dirac equation, Chen and Wang etc. \cite{16,17,18} elaborated on the fermion tunneling of black holes and black rings. It is shown that GUP may slow down the temperature of the black hole increase, and prevent the black hole from evaporating completely, which leads to the existence of the minimum non-zero mass, that is, the black hole remnant. Subsequently, the quantum tunneling radiation and remnant of G\"{o}del black hole and high dimensional Myers-Perry black hole have also been deeply studied \cite{19,20,21}. Above research proves that the black hole remnant can exist, but related issues of the stability of remnant and residue entropy have been comparatively less discussed for a Finsler black hole.

Although Einstein's general relativity described by Riemannian geometry is one of the most successful gravitational theories, it still has some problems in explaining the accelerating expansion of the universe and establishing a complete theory of quantum gravity. One has considered that the difficulties caused by general relativity may have something to do with the mathematical tools it uses. So, people try to establish a modified gravitational theory described by Finsler geometry. Finsler geometry is the most general differential geometry, which regards Riemannian geometry as its special case, and it is just Riemannian geometry without quadratic restriction \cite{22}. In recent years, the application of Finsler geometry in black hole physics has gradually aroused people's interest. People began to construct field equations of all kinds of Finsler spacetime.  With the study of Finsler field equation, people try to construct various solutions of Finsler black hole \cite{22,23,24,25}. The study of these black hole solutions makes us understand the physical properties of Finsler spacetime more deeply. In this paper, based on GUP, we take a simple Finsler black hole as an example, and investigate the remnant and residue entropy from a Rutz-Schwarzschild black hole.

\section{Tunneling radiation of a scalar particle based on GUP}

In this section, considering the effect of GUP, applying with the corrected Hamilton-Jacobi equation, we will focus on investigating the scalar particle's tunneling radiation from a Rutz-Schwarzschild black hole. Under the frame of Finsler geometry, Rutz constructed the generalized Einstein field equation and derived a Finsler black hole solution. The metric (Rutz-Schwarzschild black hole) is given by \cite{24}
\begin{eqnarray} \label{1}
ds^2  =  - \left(1 - \frac{{2M}}{r}\right)(1 - \varepsilon d\Omega /dt)dt^2  + \left(1 - \frac{{2M}}{r}\right)^{ - 1} dr^2  + r^2 d\Omega ^2.
\end{eqnarray}
Here $\varepsilon \ll 1$ is the Finsler perturbation parameter, and the line element reduces to the Schwarzschild metric when $\varepsilon= 0$. The solution is a non-Riemannian solution, and the time component of the metric depends on the tangent vector $d\Omega /dt$. It is obvious that the correction term $\varepsilon d\Omega /dt$ remains while the mass vanishes and still exists when $r\rightarrow \infty$. The metric is very different from the Schwarzschild black hole, and it is very interesting to discuss on the scalar particle's tunneling radiation based on GUP.

By taking into account the effect of GUP, in a curved spacetime, the revised Hamilton-Jacobi equation for the motion of scalar particles can be expressed as \cite{26}
\begin{eqnarray} \label{2}
g^{00} \left( {\partial _0 S + eA_0 } \right)^2  + \left[ {g^{kk} \left( {\partial _k S + eA_k } \right)^2  + m^2 } \right] \times \left\{ {1 - 2\beta \left[ {g^{jj} \left( {\partial _j S} \right)^2  + m^2 } \right]} \right\} = 0.
\end{eqnarray}
Here $\beta  = \beta _0 l_p^2 /\hbar ^2  = \beta _0 /M_p^2 c^2 $, $ \beta _0 \left( { \le 10^{34} } \right) $ is a dimensionless constant, $l_{P}$ and   $M_{P}$ are Planck length and Planck mass. Adopting rational approximation, according to the line element and the modified Hamilton-Jacobian equation, we can get the following motion equation of a scalar particle
\begin{eqnarray} \label{3}
&&- \left[ \left( 1 - \frac{2M}{r} \right)\left( 1 - \varepsilon d\Omega /dt \right) \right]^{ - 1} \left( {\partial _t S} \right)^2  - \left[ {\left( {1 - \frac{2M}{r}} \right)\left( {\partial _r S} \right)^2  + \frac{1}{{r^2 }}\left( {\partial _\theta  S} \right)^2 } \right. \nonumber \\&&+
\left. {\frac{1}{{r^2 \sin ^2 \theta }}\left( {\partial _\varphi  S} \right)^2  + m^2 } \right]\left\{ {1 - 2\beta \left[ {\left( {1 - \frac{2M}{r}} \right)\left( {\partial _r S} \right)^2 } \right.} \right. \nonumber \\&&+
\left. {\left. {\frac{1}{{r^2 }}\left( {\partial _\theta  S} \right)^2  + \frac{1}{{r^2 \sin ^2 \theta }}\left( {\partial _\varphi  S} \right)^2  + m^2 } \right]} \right\} = 0.
\end{eqnarray}
Setting $S =  - \omega t + W\left( r \right) + \Theta \left( {\theta ,\varphi } \right)$, with $\omega$ standing for the energy of a scalar particle, and inserting the action $S$ into Eq. (3), we have
\begin{eqnarray} \label{4}
&&2\beta \left( {1 - \frac{2M}{r}} \right)^2 \left( {\partial _r W(r)} \right)^4  - \left( {1 - \frac{2M}{r}} \right)\left( {\partial _r W(r)} \right)^2  \nonumber \\&&+
\left[ {\left. {\left( {1 - \frac{2M}{r}} \right)\left( {\partial _r W(r)} \right)^2  + \frac{1}{{r^2 }}\left( {\partial _\theta  \Theta (\theta ,\varphi )} \right)^2  + \frac{1}{{r^2 \sin ^2 \theta }}\left( {\partial _\varphi  \Theta (\theta ,\varphi )} \right)^2 } \right]} \right. \nonumber \\&&\times
\left[ {4\beta \left( {1 - \frac{2M}{r}} \right)}  {\left( {\partial _r W(r)} \right)^2  - 1} \right ] + \omega ^2 \left[ {\left( {1 - \frac{2M}{r}} \right)\left( {1 - \varepsilon d\Omega /dt} \right)} \right]^{ - 1}  =  - \lambda
\end{eqnarray}
and
\begin{eqnarray} \label{5}
{\rm{2}}\beta \left[ {\left. {\frac{1}{{r^2 }}\left( {\partial _\theta  \Theta (\theta ,\varphi )} \right)^2  + \frac{1}{{r^2 \sin ^2 \theta }}\left( {\partial _\varphi  \Theta (\theta ,\varphi )} \right)^2 } \right]{\rm{ = }}\lambda } \right.  ,
\end{eqnarray}
For the Eq. (5), the mode of angular momentum of a tunneling particle is associated with its components $\partial _\theta  \Theta$ and $\partial _\varphi  \Theta $ as follows \cite{27}
\begin{eqnarray} \label{6}
\frac{1}{{r^2 }}\left( {\partial _\theta  S} \right)^2  + \frac{1}{{r^2 \sin ^2 \theta }}\left( {\partial _\varphi  S} \right)^2  = L^2 .
\end{eqnarray}
Eqs. (5) and (6) yield
\begin{eqnarray} \label{7}
2\left( {L^2 } \right)^2  =\frac{\lambda}{\beta}.
\end{eqnarray}
Formula (7) shows that $\lambda$ is related to the angular momentum $L$ of the scalar particle, and it is pointed out that the angular part will have an effect on the tunneling radiation of the scalar particle.  Thus, the equation (4) can change to
\begin{eqnarray} \label{8}
&&2\beta \left( {1 - \frac{2M}{r}} \right)^2 \left( {\partial _r W(r)} \right)^4  + \left( {4m^2 \beta  + \sqrt {8\beta \lambda }  - 1} \right) \nonumber \\&&\times \left( {1 - \frac{2M}{r}} \right)\left( {\partial _r W(r)} \right)^2  + \left( {2m^2 \beta  + \sqrt {8\beta \lambda }  - 1} \right)m^2  \nonumber \\&&
+ \omega ^2 \left[ {\left( {1 - \frac{2M}{r}} \right)\left( {1 - \varepsilon d\Omega /dt} \right)} \right]^{ - 1}  - \sqrt {\frac{\lambda}{2\beta}}  + \lambda  = 0.
\end{eqnarray}
The equation (8) is solved and the higher-order term of $\beta$ is ignored, then $W(r)$ is
taken as
\begin{eqnarray} \label{9}
W(r)_ \pm   =  &&\pm \int {\frac{{\sqrt {\left( {1 - 2M/r} \right)\left( {1 - \varepsilon d\Omega /dt} \right)\left( {m^2 (1 - 2\beta m^2 ) - \lambda  + \sqrt {\lambda /2\beta} } \right) + \omega ^2 } }}{{\sqrt {\left( {1 - \varepsilon d\Omega /dt} \right)} \left( {1 - 2M/r} \right)}}} \nonumber \\&& \times
\left\{ {1 + \beta \left[ {m^2  + \omega ^2 \left[ {\left( {1 - \frac{2M}{r}} \right)\left( {1 - \varepsilon d\Omega /dt} \right)} \right]^{ - 1} } \right] +\frac{{\beta \lambda }}{2}} \right\}dr,
\end{eqnarray}
where $+ (-)$ represents the solution of the outgoing (ingoing) wave. The Laurent series is expanded at the event horizon $r=r_{+}$, and the solution of Eq. (9) is obtained by using the contour integral
\begin{eqnarray} \label{10}
W\left( {r_ +  } \right)_ \pm   =  \pm i\frac{{2\pi M\omega }}{{\sqrt {1 - \varepsilon d\Omega /dt} }}\left[ {1 + \frac{1}{2}\beta \left( {3m^2  + \lambda  + \frac{{4\omega ^2 }}{{1 - \varepsilon d\Omega /dt}}} \right) + \sqrt {\frac{{\beta \lambda }}{8}} } \,\right] + {\rm{\texttt{real part}}} . \nonumber\\
\end{eqnarray}
Applying the invariant tunneling rate under the canonical transformation \cite{28,29}, considering the influence of the time part on the tunneling rate, through the Kruskal coordinate $(T, R)$, that is $T = e^{\kappa r^* } \sinh \left( {\kappa t} \right)$ and $R = e^{\kappa r^* } \cosh \left( {\kappa t} \right)$, we get the contribution of the extra imaginary part of the time segment and have $
{\mathop{\rm Im}\nolimits} \,\,\omega t_{out\left( {in} \right)}  =  - {{\pi \omega } \mathord{\left/
{\vphantom {{\pi \omega } {2\kappa }}} \right. \kern-\nulldelimiterspace} {2\kappa }}$. As a result, the total tunneling rate of a scalar particle passing through the event horizon of Rutz-Schwarzschild black hole is as follows
\begin{eqnarray} \label{11}
 \Gamma  & \propto & \exp \left\{ {\left[ {{\mathop{\rm Im}\nolimits} \left( {\omega t_{out} } \right) + {\mathop{\rm Im}\nolimits} \left( {\omega t_{in} } \right) - {\mathop{\rm Im}\nolimits} \oint {Pdr} } \right]} \right\} \nonumber\\
 &=& \exp \left\{ { - 4\pi \frac{{2M\omega }}{{\sqrt {1 - \varepsilon d\Omega /dt} }}\left[ {1 + \frac{1}{2}\beta \left( {3m^2  + \lambda  + \frac{{4\omega ^2 }}{{1 - \varepsilon d\Omega /dt}}} \right) + \sqrt {\frac{{\beta \lambda }}{8}} } \right]} \right\}.
\end{eqnarray}
Compared with the Boltzmann factor $\Gamma  = \exp \left( -{{\omega  \mathord{\left/{\vphantom {\omega  T}} \right.\kern-\nulldelimiterspace} T}} \right)$, the corrected tunneling temperature of the black hole is
\begin{eqnarray} \label{12}
 T &=& \frac{{\sqrt {1 - \varepsilon d\Omega /dt} }}{{8\pi M}}\left[ {1 + \frac{1}{2}\beta \left( {3m^2  + \lambda  + \frac{{4\omega ^2 }}{{1 - \varepsilon d\Omega /dt}}} \right) + \sqrt {\frac{{\beta \lambda }}{8}} } \,\right]^{ - 1}  \nonumber\\
  &=& T_H \left[ {1 + \frac{1}{2}\beta \left( {3m^2  + \lambda  + \frac{{4\omega ^2 }}{{1 - \varepsilon d\Omega /dt}}} \right) + \sqrt {\frac{{\beta \lambda }}{8}} } \,\right]^{ - 1}.
\end{eqnarray}
Here $T_H  = \sqrt {1 - \varepsilon d\Omega /dt} /8\pi M $ is the Hawking temperature. According to the first law of black hole thermodynamics, the entropy of Rutz-Schwarzschild black hole is calculated as
\begin{eqnarray} \label{13}
S = \int {\frac{{dM}}{T}}  = \int {\frac{{8\pi M}}{{\sqrt {1 - \varepsilon d\Omega /dt} }}} \left[ {1 + \frac{1}{2}\beta \left( {3m^2  + \lambda  + \frac{{4\omega ^2 }}{{1 - \varepsilon d\Omega /dt}}} \right) + \sqrt {\frac{{\beta \lambda }}{8}} } \,\right ]dM.
\end{eqnarray}
From Eqs. (11)---(13), we obviously find that the scalar particle's tunneling rate, tunneling temperature and the entropy are all not only dependent on the black hole mass $M$, tunneling scalar particle's mass $m$ and energy $\omega$, but also dependent on Finsler perturbation parameter $\varepsilon$, GUP parameter $\beta$ and angular momentum parameter $\lambda$.

\section{Remnant and entropy based on thermodynamics phase transition}

On the basis of the above scalar particle's tunneling radiation, considering the GUP, now we focus on discussing on the remnant and entropy at the end of evaporation. Since all  the tunneling particles at the event horizon can be regarded as massless, the mass of scalar particles is no longer considered in the following process. According to the uncertainty relation $\Delta p \ge {\hbar  \mathord{\left/{\vphantom {\hbar  {\Delta x}}} \right.\kern-\nulldelimiterspace} {\Delta x}}$ and the lower limit of tunneling particle energy \cite{30,31} $\omega  \ge {\hbar  \mathord{\left/{\vphantom {\hbar  {\Delta x}}} \right.\kern-\nulldelimiterspace} {\Delta x}}$, near the event horizon, the uncertainty of the position can be taken as the radius of the black hole \cite{30,31}, that is $\Delta x \approx r_{BH}  = r_ +$. Consequently, the tunneling temperature of the black hole evaporating to Planck scale is
\begin{eqnarray} \label{14}
T &=& \frac{{\sqrt {1 - \varepsilon d\Omega /dt} }}{{8\pi M}}\left\{ {1 + \frac{1}{2}\beta \left( {\lambda  + \frac{{4\omega ^2 }}{{1 - \varepsilon d\Omega /dt}}} \right) + \sqrt {\frac{{\beta \lambda }}{8}} } \right\}^{ - 1}  \nonumber \\
&=& T_H \left\{ {1 - \frac{1}{2}\beta \left( {\lambda  + \frac{{4\hbar ^2 }}{{\left( {1 - \varepsilon d\Omega /dt} \right)r_ + ^2 }}} \right) - \sqrt {\frac{{\beta \lambda }}{8}} } \right\}.
\end{eqnarray}
It can be seen that the modified tunneling temperature near Planck scale is related to the properties of the background spacetime of Finsler black hole, the energy of a tunneling particle and the parameter of quantum gravitational effect. When the radius of the black hole satisfies
\begin{eqnarray} \label{15}
r_ +   <  \sqrt {\frac{{8\beta \hbar ^2 }}{{\left( {1 - \varepsilon d\Omega /dt} \right)\left( {4 - 2\beta \lambda  - \sqrt {2\beta \lambda } } \right)}}} \,   ,
\end{eqnarray}
the revised tunneling temperature $T<0$ violates the third law of thermodynamics. This means that, by considering the effect of GUP, the evaporation will stop when the tunneling temperature is infinitely close to absolute zero, which leads to a minimum radius, namely
\begin{eqnarray} \label{16}
r_i  &=& \sqrt {\frac{{8\beta \hbar ^2 }}{{\left( {1 - \varepsilon d\Omega /dt} \right)\left( {4 - 2\beta \lambda  - \sqrt {2\beta \lambda } } \right)}}}  \nonumber\\
&=& \ell _p \sqrt {\frac{{8\hbar ^2 \beta _0 }}{{\left( {1 - \varepsilon d\Omega /dt} \right)\left( {4\hbar ^2  - 2\lambda \beta _0 \ell _p^2  - \ell _p \hbar \sqrt {2\lambda \beta _0 } } \right)}}}\,   .
\end{eqnarray}
The expression (16) is represented by the mass. By using the relation between event horizon and mass $ r_{+}=2M $, near the Planck scale, the tunneling temperature can be expressed as
\begin{eqnarray} \label{17}
T = \frac{{\sqrt {1 - \varepsilon d\Omega /dt} }}{{8\pi GM}}\left\{ {1 - \frac{1}{2}\beta \left( {\lambda  + \frac{{\hbar ^2 }}{{\left( {1 - \varepsilon d\Omega /dt} \right)G^2 M^2 }}} \right) - \sqrt {\frac{{\beta \lambda }}{8}} } \right\}.
\end{eqnarray}
In order to ensure the temperature $T\geq 0$, the mass of the black hole satisfies
\begin{eqnarray} \label{18}
M \ge \sqrt {\frac{{2\beta \hbar ^2 }}{{\left( {1 - \varepsilon d\Omega /dt} \right)\left( {4 - 2\beta \lambda  - \sqrt {2\beta \lambda } } \right)G^2 }}}\,  ,
\end{eqnarray}
which implies the minimum mass of a black hole
\begin{eqnarray} \label{19}
M_{min}  &=& \sqrt {\frac{{2\beta \hbar ^2 }}{{\left( {1 - \varepsilon d\Omega /dt} \right)\left( {4 - 2\beta \lambda  - \sqrt {2\beta \lambda } } \right)G^2 }}} \nonumber\\ &=& M_p \sqrt {\frac{{2\beta _0 \hbar ^2 }}{{c^2 \left( {1 - \varepsilon d\Omega /dt} \right)\left( {4 - 2\lambda \beta _{0}/M_{p}^2 c{^2} - \sqrt {2\lambda \beta _{0}/M_{p}^2 c{^2}}}  \right)}}} \, .
\end{eqnarray}
The value is the Rutz-Schwarzschild black hole's remnant, that is $M_{res}  = M_{min}$. In order to explain the problem better, we can analyze heat capacity $C$, which is
\begin{eqnarray} \label{20}
C = T\left( {\frac{{\partial S}}{{\partial T}}} \right) = T\frac{{\partial S}}{{\partial M}}\frac{{\partial M}}{{\partial T}} = T\frac{{\partial S}}{{\partial M}}\left( {\frac{{\partial T}}{{\partial M}}} \right)^{ - 1}  = \frac{N}{F}.
\end{eqnarray}
Here $S$ is the modified black hole entropy, at the Planck scale it can be rewritten as
\begin{eqnarray} \label{21}
 S &=& \int {\frac{{dM}}{T}}  = \int {\frac{{8\pi GM}}{{\sqrt {1 - \varepsilon d\Omega /dt} }}} \left\{ {\left( {1 + \frac{1}{2}\beta \lambda  + \sqrt {\frac{{\beta \lambda }}{8}} } \right) + \left( {\frac{{\beta \hbar ^2 }}{{2\left( {1 - \varepsilon d\Omega /dt} \right)G^2 M^2 }}} \right)} \right\}dM \nonumber\\
 &=& \frac{{4\pi GM^2 }}{{\sqrt {1 - \varepsilon d\Omega /dt} }}\left( {1 + \frac{1}{2}\beta \lambda  + \sqrt {\frac{{\beta \lambda }}{8}} } \right) + \frac{{4\pi \beta \hbar ^2 }}{{G\left( {1 - \varepsilon d\Omega /dt} \right)^{{3 \mathord{\left/
 {\vphantom {3 2}} \right.
 \kern-\nulldelimiterspace} 2}} }}\ln M,
\end{eqnarray}
and the coefficients $N$ and $F$ are, respectively,
\begin{eqnarray} \label{22}
N = \left[ {2G^2 M^2 \left( {1 - \varepsilon d\Omega /dt} \right)\left( {1 - \frac{1}{2}\beta \lambda  - \sqrt {\frac{{\beta \lambda }}{8}} } \right) - \beta \hbar ^2 } \right] \times  \left[ {16\pi G^3 M^4 \sqrt {1 - \varepsilon d\Omega /dt} } \right]
\end{eqnarray}
and
\begin{eqnarray} \label{23}
 F = &&\left[ {2G^2 M^2 \left( {1 - \varepsilon d\Omega /dt} \right)\left( {1 + \frac{1}{2}\beta \lambda  + \sqrt {\frac{{\beta \lambda }}{8}} } \right) + \beta \hbar ^2 } \right]   \nonumber \\ && \times
\left[ {3\beta \hbar ^2  - 2G^2 M^2 \left( {1 - \frac{1}{2}\beta \lambda  - \sqrt {\frac{{\beta \lambda }}{8}} } \right)\left( {1 - \varepsilon d\Omega /dt} \right)} \right].
\end{eqnarray}
From the expression (20), we find
\begin{eqnarray} \label{24}
M = M_{Cr}  = M_p \sqrt {\frac{{6\beta _0 \hbar ^2 }}{{c^2 \left( {1 - \varepsilon d\Omega /dt} \right)\left( {4 - 2\lambda \beta _{0}/M_{p}^2 c{^2} - \sqrt {2\lambda \beta _{0}/M_{p}^2 c{^2}}} \right)}}}\, ,\,\,\,  C\to 0 \,  ,
\end{eqnarray}
\begin{eqnarray} \label{25}
M \to M_{min}  = M_{res}  = M_p \sqrt {\frac{{2\beta _0 \hbar ^2 }}{{c^2 \left( {1 - \varepsilon d\Omega /dt} \right)\left({4 - 2\lambda \beta _{0}/M_{p}^2 c{^2} - \sqrt {2\lambda \beta _{0}/M_{p}^2 c{^2}}} \right)}}},\,\,\,  C\to 0 \,   .\nonumber\\
\end{eqnarray}
The expression (25) is consistent with the expression (19). From above equations we can see that, for the Rutz-Schwarzschild black hole, considering the tunneling of scalar particles, when the black hole evaporates to the Planck scale, the phase transition takes place, leaving a stable remnant. Then, based on the expressions (14), (17), (20) and (21), we draw the following thermodynamic curves 1---5, and further analyze the entropy and remnant in detail.  We set $\gamma  = \varepsilon d\Omega /dt$ and $M_p  = c = \hbar  = 1$ in all of the following drawings for research convenience.

As Figure1 and Figure 2 show, in a large range of radius and mass, for the cases $\beta=0$ and $\beta=1$, the tunneling temperatures tend to be consistent. But as the radius (or mass) closes to the critical value $r_{cr}$ (or $M_{cr}$), these curves become markedly different. The tunneling temperature reaches the maximum value at $r_{cr}$ (or $M_{cr}$) for the case of $\beta=1$, then it decreases with the decrease of radius (or mass). Finally, the tunneling temperature tends to zero when the radius (or mass) attains to the minimum value $r_{i}$ (or $M_{res}$) for the case of $\beta=1$, which leads to remnant. We can also explain it through the Figure 3.

\begin{figure}[H]
\centering
\includegraphics[width=0.6\textwidth]{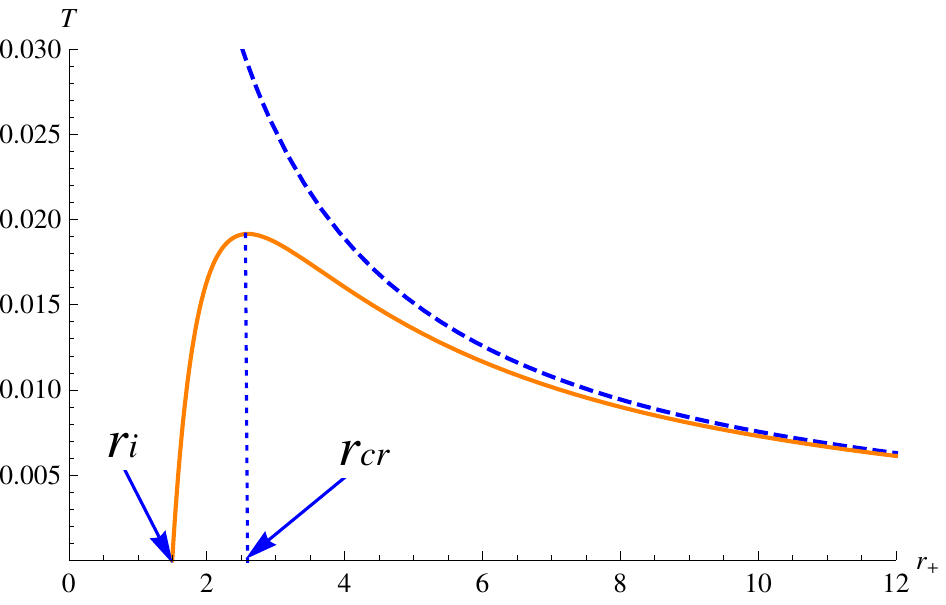}
\caption{\small The figure shows the temperature $T$ versus the radius of event horizon $r_{+}$ for varying $\beta$. Above blue dashed curve is $\gamma=0.1$ and $\beta=0$. Below orange curve is $\gamma=0.1$ and $\beta=1$.}
\end{figure}

\begin{figure}[H]
\centering
\includegraphics[width=0.6\textwidth]{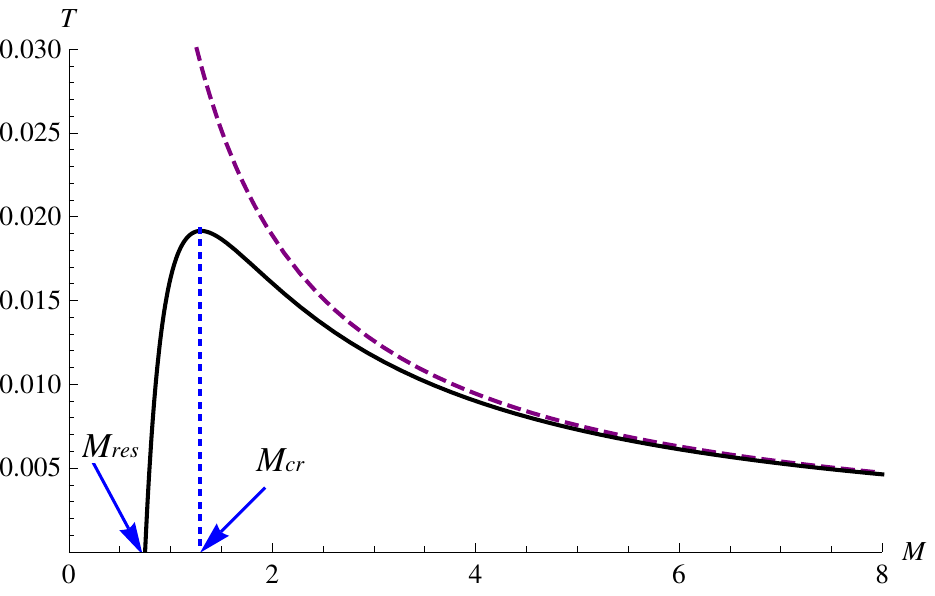}
\caption{\small The figure shows the temperature $T$ versus the mass $M$ for varying $\beta$. Above purple dashed curve is $\gamma=0.1$ and $\beta=0$. Below black curve is $\gamma=0.1$ and $\beta=1$.}
\end{figure}

In Figure 3, the heat capacity monotonically increases with the mass increasing for the case of $\beta=0$, and it is always negative. However, for the case of $\beta=1$, we find that, at $M=M_{cr}$, heat capacity is divergent, which means the existence of phase transition. That is the black hole undergoes a phase transition from $C<0$ (unstable hole) to $C>0$ (stable hole), which also corresponds to the phase transitions in Figure1 and Figure 2 at the critical value $r_{cr}$ and $M_{cr}$. Furthermore, we can see that the mass tends to minimum value (namely $M_{min}=M_{res}$) when $C\rightarrow 0$. In addition, we also note that the same problem is treated in noncommutative geometry in Ref. [32] (see Figure 7). The result shows that the heat capacity tends to zero when the mass approaches to a certain value, but some unusual feature is also presented.

\begin{figure}[H]
\centering
\includegraphics[width=0.6\textwidth]{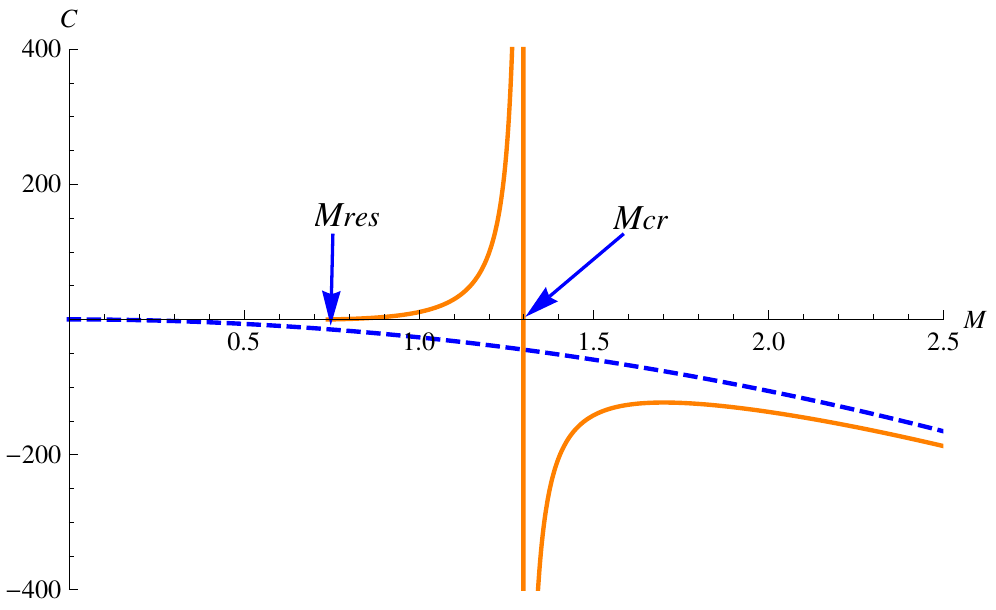}
\caption{\small The figure shows the heat capacity $C$ versus the mass $M$ for varying $\beta$. The blue dashed curve is $\gamma=0.1$ and $\beta=0$. The orange curve is $\gamma=0.1$ and $\beta=1$.}
\end{figure}

In Figure 4 and Figure 5, the crucial difference is shown for the two curves. To clearly see the difference between them, we can refer to Figure 5. The traditional entropy is zero as the mass becomes zero for the case of $\beta=0$. However, by considering GUP, the entropy is also revised for the case of $\beta=1$, and there exits minimum $S_{min}$ (namely residual entropy $S_{res}=S_{min}$) at the remnant, which means the entropy is no longer zero in the final stages of black hole evaporation.

\begin{figure}[H]
\centering
\includegraphics[width=0.6\textwidth]{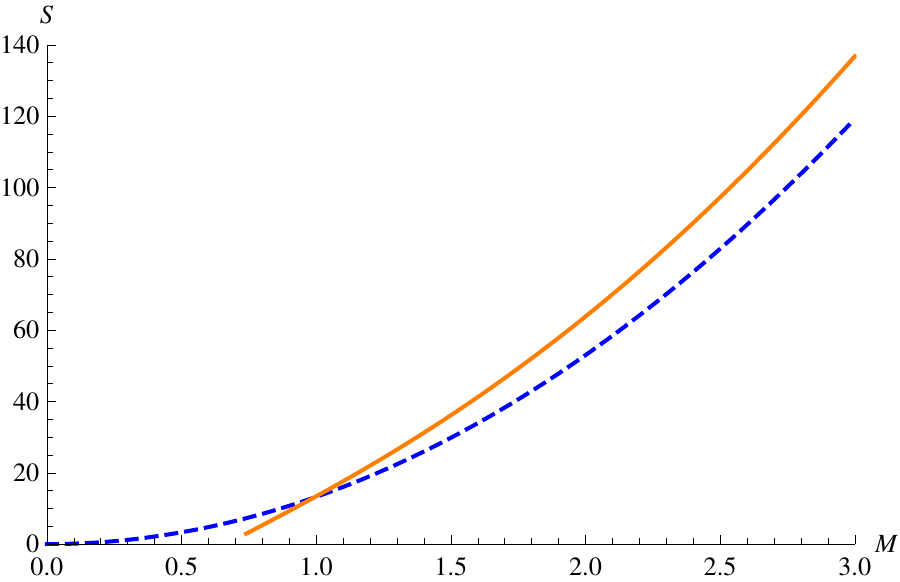}
\caption{\small The figure shows the entropy $S$ versus the mass $M$ for varying $\beta$ on a large scale of entropy. The blue dashed curve is $\gamma=0.1$ and $\beta=0$. The orange curve is $\gamma=0.1$ and $\beta=1$.}
\end{figure}

\begin{figure}[H]
\centering
\includegraphics[width=0.6\textwidth]{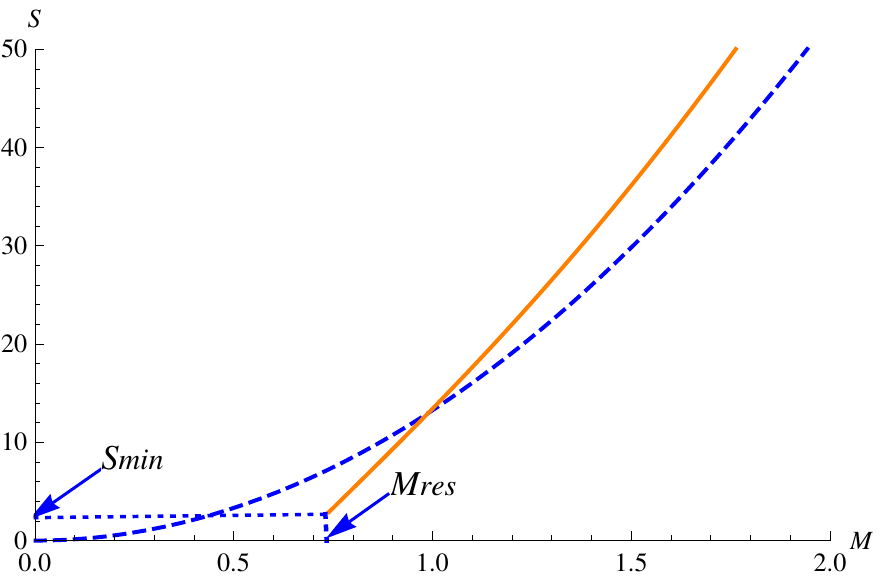}
\caption{\small The figure shows the entropy $S$ versus the mass $M$ for varying $\beta$ within a small scale of entropy. The blue dashed curve is   $\gamma=0.1$ and $\beta=0$. The orange curve is $\gamma=0.1$ and $\beta=1$.}
\end{figure}

To see clearly the Finsler perturbation parameter and angular momentum parameter effect on remnant and residual entropy, we present the following tables for different $\gamma$ and $\lambda$ when $\beta=1$. From Tables 1 and Tables 2, we can find that remnant and residual entropy increase as Finsler perturbation parameter increase, and they are also increase as angular momentum parameter increase.

\begin{table}
\centering \caption{Remnant and residual entropy with $\lambda=0.001$ for different $\gamma$} \label{parset}
\begin{tabular*}{\columnwidth}{@{\extracolsep{\fill}}llll@{}}
\hline
 \multicolumn{1}{c}{$\gamma$} & \multicolumn{1}{c}{$M_{res}$} & \multicolumn{1}{c}{$S_{res}$} \\
\hline
 0.1  & 0.749748       & 3.29388         \\
 0.01     & 0.714856        & 1.90766         \\
 0.0001     & 0.711309        & 1.76666        \\
\hline
\end{tabular*}
\end{table}

\begin{table}
\centering \caption{Remnant and residual entropy with $\gamma=0.1$ for different $\lambda$} \label{parset}
\begin{tabular*}{\columnwidth}{@{\extracolsep{\fill}}llll@{}}
\hline
 \multicolumn{1}{c}{$\lambda$} & \multicolumn{1}{c}{$M_{res}$} & \multicolumn{1}{c}{$S_{res}$} \\
\hline
 0.5  & 1.054090       & 22.8521         \\
 0.1     & 0.814124        & 7.17324         \\
 0.01     & 0.760867        & 3.95553         \\
\hline
\end{tabular*}
\end{table}

In addition , when $\beta=0$ and $\varepsilon=0$ (namely $\gamma=0$), the results reduce to that of Schwarzschild black hole without GUP . In this case, the black hole can radiate continuously until the mass and entropy decrease to zero. In the case of $\beta=1$ and $\varepsilon=0$, the results reduce to that of Schwarzschild black hole with GUP. From Eq. (21) and (25), we can see that, the remnant and the residue entropy in the frame of Finsler geometry are larger than those of Schwarzschild black hole. Moreover , we also see that the influence of Finsler parameter on remnant and residue entropy from Table 1.

\section{Discussion on remnant based on Parikh-Wilczek tunneling}
Based on GUP, using Parikh-Wilczek tunneling method, we continue to explore the quantum tunneling and remnant of a Rutz-Schwarzschild black hole in the frame of Finsler geometry. The Rutz-Schwarzschild metric in the Painleve coordinate system can be rewritten as
\begin{eqnarray} \label{26}
ds^2  =  - \left(1 - \frac{{2M}}{r}\right)(1 - \varepsilon d\Omega /dt)dt'^2  +2\sqrt{\frac{{2M}}{r}(1 - \varepsilon d\Omega /dt)}dt'dr+  dr^2  + r^2 d\Omega ^2,
\end{eqnarray}
which is obtained by the coordinate transformation
\begin{eqnarray} \label{27}
dt\rightarrow dt'-\frac{\sqrt{2M/r}}{1-2M\sqrt{1 - \varepsilon d\Omega /dt}/r}dr.
\end{eqnarray}
We can derive the radial null geodesics
\begin{eqnarray} \label{28}
\dot{r}\equiv \frac{dr}{dt'}=\sqrt{1 - \varepsilon d\Omega /dt}\left(\pm1-\sqrt{\frac{2M}{r}}\,\right),
\end{eqnarray}
where $+ (-)$ represents the solution of the outgoing (ingoing) geodesics. Now, we consider the influence of GUP with a minimal length on quantum tunneling. The expression GUP with a minimal length is \cite{12}
\begin{eqnarray} \label{29}
\Delta x \Delta p\geq \hbar \left[1+\beta(\Delta p)^{2}\right],
\end{eqnarray}
and the generalized energy \cite{11} is
\begin{eqnarray} \label{30}
\omega=E(1+\beta E^{2}),
\end{eqnarray}
where $\beta=\alpha^{2}l_{p}^{2}$. From the GUP expression, the revised commutation relation between the radial coordinate and the conjugate momentum becomes
\begin{eqnarray} \label{31}
[r,p_{r}]=i\hbar(1+\beta p_{r}^{2}).
\end{eqnarray}
In the classical limit, above commutation relation can be replaced by the following poisson bracket
\begin{eqnarray} \label{32}
\left\{r,p_{r}\right\}=1+\beta p_{r}^{2}.
\end{eqnarray}
In order to obtain the imaginary part of the action, we use the following deformed Hamiltonian equation
\begin{eqnarray} \label{33}
\dot{r}=\left\{r,H\right\}=\left\{r,p_{r}\right\}\left.\frac{dH}{dp_{r}}\right|_{r}.
\end{eqnarray}
In the framework of Parikh and Wilczek's tunneling radiation, the imaginary part of the action can be given by
\begin{eqnarray} \label{34}
Im I=Im \int_{r_{in}}^{r_{out}}p_{r}dr=Im \int_{r_{in}}^{r_{out}} \int_{0}^{p_{r}} p'_{r}dr.
\end{eqnarray}
According to the work in Ref. [12], the Hamiltonian is $H=M-\omega'$, $p_{r}^{2}\approx \omega'^{2}$, $p_{r}\approx \omega'$.
Thus, by eliminating the momentum in the favor of the energy in Eq. (34), we have
\begin{eqnarray} \label{35}
{\mathop{\rm Im}\nolimits} I &=& {\mathop{\rm Im}\nolimits} \int_M^{M - \omega } {\int_{r_{in} }^{r_{out} } {\frac{{1 + \beta \omega '^2 }}{{\dot r}}} } drdH'  \nonumber\\ &=& {\mathop{\rm Im}\nolimits} \int_M^{M - \omega } {\int_{r_{in} }^{r_{out} } {\frac{{1 + \beta \omega '^2 }}{{\left( {1 - \varepsilon d\Omega /dt} \right)\left( {1 - \frac{{2(M - \omega ')}}{r}} \right)}}} } drd(M - \omega ')
  \nonumber\\ &=& \frac{{4\pi M\omega }}{{\left( {1 - \varepsilon d\Omega /dt} \right)}}\left[ {1 - \frac{\omega }{{2M}} + \beta \omega ^2 \left( {\frac{1}{3} - \frac{\omega }{{4M}}} \right)} \right].
\end{eqnarray}
The tunneling rate with effect of GUP with a minimal length is therefore
\begin{eqnarray} \label{36}
 \Gamma  \sim \exp \left( { - 2{\mathop{\rm Im}\nolimits} S} \right) = \exp \left[ { - \frac{{8\pi ME}}{{\left( {1 - \varepsilon d\Omega /dt} \right)}} + \frac{{4\pi E^2 }}{{\left( {1 - \varepsilon d\Omega /dt} \right)}} - \frac{{32\pi ME^3 }}{{3\left( {1 - \varepsilon d\Omega /dt} \right)}}} \right. \nonumber\\
 \left. {\begin{array}{*{20}c}
   {} & {}  \\
\end{array} + \frac{{10\pi \beta E^4 }}{{\left( {1 - \varepsilon d\Omega /dt} \right)}} + {\rm O}\left( {\beta ^2 } \right)} \right] = \exp (\Delta S), \end{eqnarray}
where $\Delta S$ is the difference in black hole entropies before and after emission \cite{11,12,33}. It is shown that Parikh-Wilczek's procedure is also valid for the Rutz-Schwarzschild metric. When $\beta=0$ and $\varepsilon=0$, the result reduces to the result of Schwarzschild black hole, which is the same with that derived by Parikh and Wilczek \cite{33}. Note that, for a Schwarzschild black hole, with Parikh-Wilczek tunneling mechanism, the quantum tunneling of massless particles from black hole horizon has been studied based on GUP \cite{11,12}. It is obvious that our result coincides with the one in Ref. [12] when $\varepsilon=0$. We find that the above equation (36) gives tunneling rate's correction which caused by the varied background spacetime and GUP.  When $\beta=0$, the tunneling temperature
\begin{eqnarray} \label{37}
T = \left(1 + \frac{E}{{2M}}\right)T_H
\end{eqnarray}
is higher than the original temperature $T_H$.  This predicts the varied spacetime can accelerate the evaporation and the remnant does not arise. when GUP is taken into account, the tunneling temperature
\begin{eqnarray} \label{38}
T = \left(1 + \frac{E}{{2M}} - \frac{4}{3}\beta E^2  + \frac{5}{{4M}}\beta E^3 \right)T_H
\end{eqnarray}
is lower than the original Hawking temperature $T_H$,  which implies that GUP can slow down the increase of the Hawking temperature and stop the black hole from completely evaporating. As a result, a remnant of the black hole is left at the end of evaporation.

\section{Conclusion}

In conclusion, employing the tunneling radiation of a scalar particle and generalized uncertainty principle, we study the remnant and residue entropy from a Rutz-Schwarzschild black hole based on black hole thermodynamic.  Firstly, based on the Hamilton-Jacobi equation revised by GUP, we present the modified tunneling temperature-uncertainty relation and modified entropy-uncertainty relation by using quantum tunneling method. Then, using the generalized uncertainty principle, we calculate the remnant and residual entropy after evaporation when the black hole reaches the Planck scale.  Finally, based on black hole thermodynamic phase transition, a detailed analysis of whether there is a stable remnant and residual entropy in the final stage of evaporation is given, that is, the thermodynamic stability of a black hole is related to its thermal capacity and temperature. In order to ensure the thermal stability of a black hole, when the black hole evaporates to the Planck scale, the black hole with negative heat capacity must be transformed into a black hole with positive heat capacity. As a result, at remnant, the modified tunneling temperature  and heat capacity tend to zero, and modified entropy reaches the minimum value, which imply the black holes are in thermal equilibrium with the outside environment. In addition, the effects of the Finsler perturbation parameter and angular momentum parameter on remnant and residual entropy are also discussed.

The emergence of Finsler black hole solution infuses new vitality into general relativity and puts forward a new research and development idea for black hole physics theory. In this paper, we consider a simple Finsler black hole. For the Rutz-Schwarzschild black hole, we investigate the remnant and residue entropy based on the scalar particles' tunneling radiation via semi-classical Hamilton-Jacobi method. In view of the size of the Finsler perturbation parameter $\varepsilon < 3.15 \times 10^{ - 4} s$ in geometrical units \cite{24}, the small physical quantity is ignored and we approximately adopt Hamilton-Jacobi equation based on GUP to study tunneling radiation from horizon. Strictly speaking, we should apply new Hamilton-Jacobi equation in the framework of complicated Finsler geometry to discuss the related problems. Up to now, Finsler gravity theory has not been established completely. Here, adopting the approximate method we only to understand simply some quantum results of Finsler gravity. We hope that our research can provide some enlightenments for the thermodynamic evolution of black hole physics.

We also note that, the expression of GUP is not unique, which gives rise to different correction on tunneling radiation. Therefore, the remnant and residue entropy should also be distinct from that in the frameworks of other GUPS. We are going to discuss the related issues in further.

\section*{Acknowledgements}

This work is supported in part by the National Natural Science Foundation of China (Grant No. 11703018) and Natural Science
Foundation of Liaoning Province, China (Grant No. 20180550275).

\end{document}